# Distribution of defect clusters in the primary damage of ion irradiated 3C-SiC

C. Liu[a], I. Szlufarska[a,b,*]

[a] University of Wisconsin-Madison, Department of Engineering Physics, 1500 Engineering Dr., Madison, WI, 53706, U.S.A.

[b] University of Wisconsin-Madison, Department of Material Science and Engineering, 1509 University Ave., Madison, WI, 53706, U.S.A.

* Corresponding author: szlufarska@wisc.edu

## Abstract

We report a statistical analysis of sizes and compositions of clusters produced in cascades during irradiation of SiC. The results are obtained using molecular dynamics (MD) simulations of cascades caused by primary knock-on atoms (PKAs) with energies between 10 eV and 50 keV. The results are averaged over six crystallographic directions of the PKA and integrated over PKA energy spectrum derived from the Stopping and Range of Ions in Matter (SRIM) code. Specific results are presented for 1 MeV Kr ion as an example of an impinging particle. We find that distributions of cluster size $n$ for both vacancies and interstitials obey a power law $f = A/n^S$ and these clusters are dominated by carbons defects. The fitted values of $A$ and $S$ are different than values reported for metals, which can be explained through different defect energetics and cascade morphology between the two classes of materials. In SiC, the average carbon ratio for interstitial clusters is 91.5%, which is higher than the ratio of C in vacancy clusters, which is 85.3%. Size and composition distribution of in-cascade clusters provide a critical input for long-term defect evolution models.

# 1. Introduction

Silicon carbide (SiC) has been proposed for use in many nuclear reactors applications, such as fuel coating of high temperature gas-cooled reactor , structural components in fission nuclear reactors, and blanket structures for fusion energy systems [1]. In reactor environments materials are exposed to irradiation, which can lead to undesirable changes, such as swelling and creep. These mechanical properties depend on the specific nature of defects and defect clusters in the material [2–4]. Point defects in SiC have been studied extensively and are relatively well understood [5,6]. A few studies of stable defect clusters [5,7–10] and their kinetics [11–14] in irradiated SiC have also been reported. Long time evolution of clusters in irradiated 3C-SiC has been captured in kinetic Monte Carlo [15], rate theory [14,16], and cluster dynamics [17] models. However, these studies did not address the issue of what clusters form directly during cascade. This issue has been investigated in other materials, including W, Fe, Be, Zr, Mo [18–22] and ferritic steels[23]. In these references, size distribution of clusters (where cluster size is defined as the number of defects in a cluster) generated by a cascade follows generally some type of a power law with different coefficients. However, a recent study in tungsten [24] predicts that the power law behavior may break down when clusters size is significantly large. In SiC, formation of small clusters directly in cascades has been suggested to be necessary in order to match predictions of clusters dynamics model to experimentally observed cluster size distributions [17]. Distribution of intra-cascade clusters provides initial conditions for model of long-term defect evolution and it is therefore important to determine both the functional form and the quantitative coefficient for such distribution.

A few molecular dynamics (MD) simulations have been performed to address the question of what clusters form in SiC directly during irradiation. However, these studies have been performed only for limited conditions (e.g., one primary knock-on (PKA) energy and direction). For example, Devanathan *et al.*[25] analyzed cascades caused by a 10 keV Si PKA along the [4 11 $\overline{95}$] direction and Gao *et al.*[26] analyzed three 50 keV Si PKAs along [1 3 5] direction. In addition, Weber *et al.*[27], Devanathan *et al.*[28] and Gao *et al.*[29] simulated cascades generated by PKAs along different directions with 8 energies selected from 0.25 keV to 50 keV. These authors reported clusters consisting of only up to 4 defects and size distribution of interstitial clusters were only reported for PKA energies of 10 keV and 50 keV. One should also note that PKAs with energies lower than 0.25 keV account for approximately 60% of all PKAs (as will be shown later); however, there are few investigations on cascades caused by such PKA in this energy range. There is also the question of the criterion used to define when point defects form a cluster, which will be further addressed in Section 3.2.1. Even less information has been available regarding statistics of clusters in terms of their composition. One previous study based on genetic algorithms combined with density functional theory (DFT) calculations [30] revealed that composition of thermodynamically stable clusters depends on the cluster size ($n$). Specifically, it was found that Si-only clusters are likely unstable, and clusters are predominantly C-only for $n \leq 10$ and stoichiometric for $n > 10$. However, this study did not provide information on compositions of clusters formed directly during cascade, which is a non-equilibrium process.

In order to obtain a statistical description of cluster sizes and their composition in a primary radiation damage, it is important to analyze cascades caused by different PKAs as well as the energy spectrum of PKAs caused by specific ions or neutrons. Below, we report the results of MD simulations of displacement cascades where we determined how distribution of debris from PKAs varies with PKA species, energies, and directions. We have also applied Stopping and Range of Ions in Matter (SRIM) simulation code to determine PKAs species and energies when SiC is irradiated with 1 MeV Kr ion as a representative example. Finally, we combine statistical results from MD and SRIM to determine the total distribution of defect clusters (i.e., their size, type, and composition) generated by one 1 MeV Kr ion. Our results can be generalized to irradiation with other ions and energies by combining results of our MD simulations with new SRIM calculations, following the procedure outlined in this paper.

## 2. Simulation methods
## 2.1 SRIM calculations

The SRIM software was applied to calculate PKA energy distribution with full damage cascades model for cubic (β) silicon carbide (SiC) irradiated with 1 MeV Kr ions. In order to obtain statistically relevant information, 2,000 Kr ion are used in the SRIM simulation. Threshold displacement energies for C and Si atoms are 20 eV [28,31,32]and 35 eV[28,32,33], respectively. PKAs have two ways to dissipate their initial kinetic energies, $E_{\text{PKA}}$, which are elastic interactions with other atoms and electron excitations. The fraction of PKA's energy that is dissipated by transferring to other nuclei and thereby generating atomic displacements is called the damage energy, $E_{\text{d}}$, which is a function of $E_{\text{PKA}}$. The ratios between $E_{\text{d}}$ and $E_{\text{PKA}}$ for C and Si atoms can be calculated based on the model developed by Norgett *et al.* [34] and are shown in Figure 1. It can be seen that as the energy of PKA increases, the ratio of the damage energy and the PKA energy decreases, and the fraction of PKA's kinetic energy, $E_{\text{PKA}}$, dissipated through electron excitation increases [35]. The above ratio decreases faster for C PKAs than for Si PKAs, which is due to the lower atomic mass of C as compared to the atomic mass of Si. For the same $E_{PKA}$, C has a higher velocity and dissipates more energy through electron excitation.

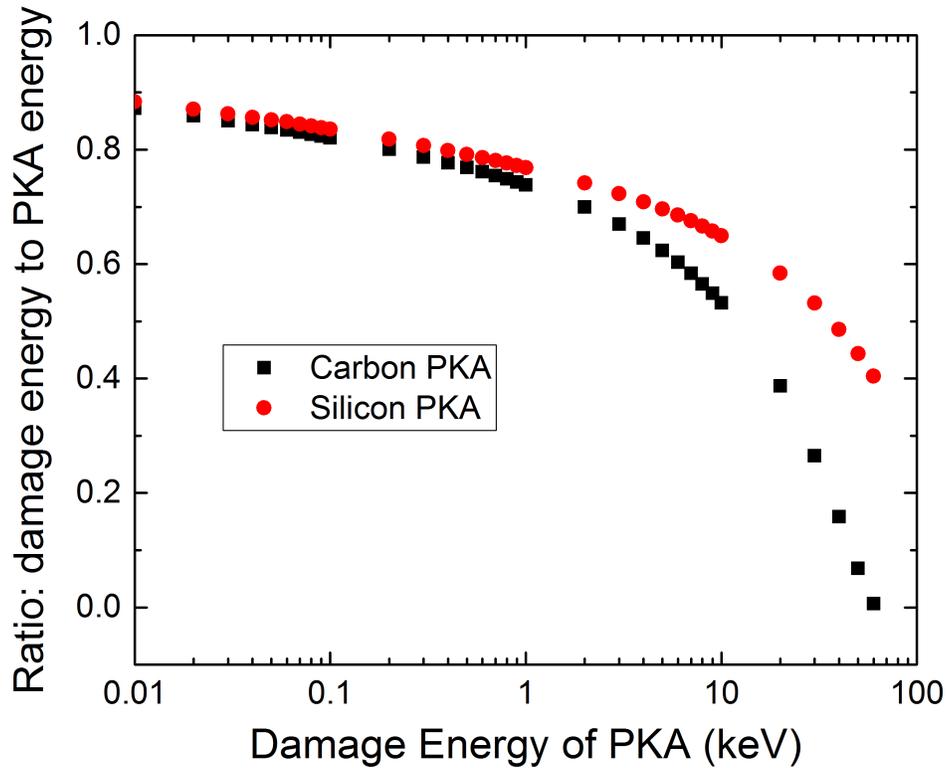

**Figure 1** Ratio between damage energy $E_d$ and PKA energy $E_{PKA}$ as a function of PKA damage energy, which is calculated based on the model developed by Norgett *et al.*[34]

## 2.2 Cascade simulations

Cascade simulations require a repulsive interaction at very close interatomic distances to better represent ballistic and thermal processes that occur during ion bombardment [36]. There are a number of potentials that have been used in MD simulations of cascades in SiC. Among the most popular potentials are Tersoff/ZBL [37], Tersoff potential modified by Devanathan *et al.* [25], and Gao-Weber potential [38]. The effect of the potential choice on the number of point defects generated during cascades is shown in Table 1, using Si 10 keV PKA cascade as an example. The number of defects generated with Tersoff/ZBL potential lies between the values predicted by the other two potentials and therefore we choose the Tersoff/ZBL potential for the current study. This potential has been widely used in literature to study radiation effects in SiC [36,39–43]. Simulations follow the same procedure as discussed in detail in Ref. [36] and are carried out using the Large-scale Atomic/Molecular Massively Parallel Simulator (LAMMPS) [44] package.

**Table 1**. Number of point defects generated in MD simulations of cascades with 10 keV Si PKA at 300K by three different SiC potentials: Tersoff potential modified by Devanathan et al. [25], Tersoff/ZBL [37] and Gao-Weber potential [38]. $V_C$ and $V_{Si}$ stand for C and Si vacancies, $I_C$ and $I_{Si}$ represent C and Si interstitials, $C_{Si}$ is a C antisite (C on Si sublattice), and $Si_C$ is a Si antisite (I on C sublattice).

| Point Defects | Tersoff-Devanathan | Tersoff/ZBL | Gao-Weber |
| --- | --- | --- | --- |
| $V_C$ | 60 | 98.6 | 120 |
| $V_{Si}$ | 16 | 22.5 | 78 |
| $I_C$ | 60 | 112.2 | 131 |
| $I_{Si}$ | 16 | 9 | 70 |
| $C_{Si}$ | 4.5 | 18.3 | 8.5 |
| $Si_C$ | 3 | 31.8 | 15 |
| Total | 159.5 | 292.5 | 422.5 |

The total numbers of atoms in simulations corresponding to each damage energy (or each PKA energy) are listed in Table 2. For damage energy smaller than 1 keV, the cascade simulation box consists of 24×24×22 conventional unit cells. Langevin thermostat is applied to 8.62 Å -thick layers on five sides of the simulation box. C/Si PKAs are initiated from locations near to the sixth face, where no thermostat is maintained. For damage energies equal to 1 keV and 2 keV, the simulation box consists of 48×48×44 unit cells and the thermostat layers are 17.24 Å thick. For damage energy larger than 2 keV and smaller than 10 keV, the number of unit cells in the simulation box varies from 48×48×44 to 108×108×104, which sizes were chosen to ensure that cascade debris do not cross the thermostat layers. For simulation larger than 10 keV, the simulation boxes are 128×128×124 and in this case we do not use a thermostat. The reasons are as follows: the purpose of adding thermostat is to dissipate energy induced by the PKA and to maintain the simulation box at a constant temperature. When simulation box is sufficiently large, the increase in temperature caused by a PKA is negligible and so is the effect of adding a thermostat. On the other hand, due to the uncertainty of cascade debris geometry, interaction between cascade debris with thermostat may be difficult to avoid and it is safer not to include the thermostat in the simulations. We have tested the effect of including a thermostat layer on the example of 10 keV Si PKA and found that the difference in the total production of point defects (PDs) from the case when there is no thermostat is smaller than 1%.

**Table 2**. Parameters used in MD simulations of displacement cascades.

| Carbon PKA Energy $E_{\text{PKA}}(keV)^a$ | Silicon PKA Energy $E_{\text{PKA}}(keV)^a$ | Damage energy $E_d(keV)^{b,c}$ | Carbon energy ratio $R_{\text{C}}{}^d$ | Silicon energy ratio $R_{\text{Si}}{}^d$ | Number of atoms in simulation |
|---|---|---|---|---|---|
| 0.012 | 0.011 | 0.01 | 0.87 | 0.88 | 101,376 |
| 0.024 | 0.023 | 0.02 | 0.86 | 0.87 | 101,376 |
| 0.060 | 0.059 | 0.05 | 0.84 | 0.85 | 101,376 |
| 0.123 | 0.120 | 0.1 | 0.82 | 0.84 | 101,376 |
| 0.25 | 0.245 | 0.2 | 0.80 | 0.82 | 101,376 |
| 0.65 | 0.631 | 0.5 | 0.77 | 0.79 | 101,376 |
| 1.35 | 1.30 | 1 | 0.74 | 0.77 | 811,008 |
| 2.86 | 2.70 | 2 | 0.70 | 0.74 | 811,008 |
| 8.01 | 7.19 | 5 | 0.62 | 0.70 | 1,722,368 |
| 18.8 | 15.4 | 10 | 0.53 | 0.65 | 5,203,968 |
| 51.6 | 34.3 | 20 | 0.39 | 0.58 | 6,443,008 |
| 730 | 113 | 50 | 0.07 | 0.44 | 16,252,928 |

[a] Energy of PKA which is gained during a collision with impinging ion.

[b] Damage energy of PKA calculated based on the NRT theory.

[c] Damage energies of PKA in current simulations include 0.01, 0.02, 0.03, ……, 0.09, 0.1, 0.2, 0.3, ……, 0.9, 1.0, 2.0, 3.0, ……, 9.0, 10.0, 20.0, 30.0, ……, 50.0 keVs. For simplicity, only several of them are listed in this table.

[d] $R_{\text{C/Si}} = E_d/E_{PKA}$ the ratio between the damage energy and the PKA energy.

Periodic boundary conditions are applied in all three spatial directions of the simulation box. PKAs were given an instantaneous velocity corresponding to specific damage energies and we performed multiple simulations where the velocity vector was aligned with six directions, i.e., $[00\bar{1}]$, $[01\bar{1}]$, $[11\bar{1}]$, $[1\bar{1}\bar{1}]$, $[13\bar{3}]$, and $[13\bar{5}]$. PKA damage energies are 0.01, 0.02, 0.03, … , 0.09, 0.1, 0.2, 0.3, … , 0.9, 1.0, 2.0, 3.0,…, 9.0, 10.0, 20.0, 30.0, …., 50.0 keV. For simplicity, only several damage energies and their corresponding simulation details are shown in Table 2.

To simplify our simulations, we build on the conclusions of Devanathan *et al.* [28] who conducted MD simulations of cascades with both C and Si PKAs with energies 0.25 keV, 0.5 keV, 1.0 keV, 2.5 keV, and 5 keV. These authors have shown that the difference in the production of PDs between C and Si PKAs decreases as the PKA damage energy increases. To further test this observation, we have performed simulations with 10 keV PKAs aligned along six different directions, using both C and Si as PKAs. We found that the difference in PD production (averaged over all directions) is around 8% (see Table 3). Consequently, we assume that cascades caused by C PKAs with damage energies larger than 5 keV can be represented by cascade caused by Si PKA with the same damage energies.

**Table 3**. Number of all point defects generated by 10 keV Si and C PKAs along different crystallographic directions. The last row shows a difference between the defects generated with Si and C PKAs.

| PKA | $[00\bar{1}]$ | $[01\bar{1}]$ | $[11\bar{1}]$ | $[1\bar{1}\bar{1}]$ | $[13\bar{3}]$ | $[13\bar{5}]$ | Average |
|---|---|---|---|---|---|---|---|
| Si | 309 | 313 | 286 | 284 | 291 | 272 | 292.5 |
| C | 274 | 260 | 274 | 278 | 268 | 256 | 268.3 |
| Difference | 11.3% | 16.9% | 4.2% | 2.1% | 7.9% | 5.8% | 8.3% |

## 2.3 Integration of PKA energy spectrum and MD cascades

PDs generated in MD cascade simulations are analyzed using the Voronoi cell method, which is described in Ref. [36]. There are six kinds of PDs in SiC - carbon interstitial ($C_I$), silicon interstitial ($Si_I$), carbon vacancy ($V_C$), silicon vacancy ($V_{Si}$), carbon antisite ($C_{Si}$), and silicon antisite ($Si_C$),). Interstitial and vacancy clusters are defined based on a cut-off distance, i.e., two PDs are identified as clustered if the distance between them is smaller than the second nearest neighbor (NN), 3.11 Å. For example, if there are three PDs, namely A, B, and C. If the distance between A and B are less than 3.11 Å, they are identified as a cluster of size 2. If the distance between A and C, or/and, B and C is less than 3.11 Å, and no other PDs are within 3.11 Å from any of them, these three PDs form a cluster of size 3. In this case, A and B are not identified as a cluster of size 2. All clusters are further analyzed in terms of their size, number, and composition. Point defect and clusters caused by PKAs with the same species and the same energies but different directions are averaged over PKA directions. This analysis yields cluster information for very specific conditions, i.e., a specific PKA and PKA energy. However, fast ions impinging on a material generate multiple PKAs with a range of energies. In order to obtain the total distribution of clusters generated by fast ions during irradiation, we combine information from cascades generated by each PKA (obtained using MD simulations) and the PKA energy distribution (calculated from SRIM).

## 3. Results and Discussion
### 3.1 PKA Spectrum

SRIM simulations of 2,000 1MeV Kr ions impinging on SiC generate 326,634 carbon and 295,925 silicon PKAs. The energy spectrum of these PKAs is shown in Figure 2 (a), which reveals that the number of PKAs decreases as the damage energy increases and more than 83.8% PKAs have energies in the range between 0 and 1 keV. The entire energy range of PKAs is large and spans from several eV's to hundred keV's.

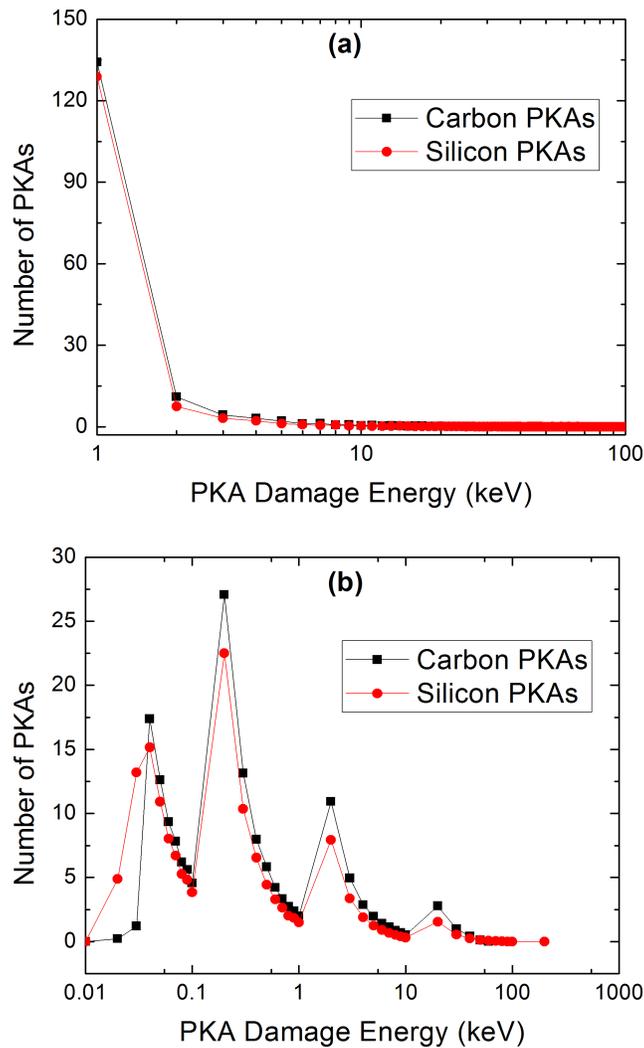

**Figure 2** Number of PKAs in each energy interval generated by one 1 MeV Kr ion in 3C-SiC. In (a), energy interval equals to 0.01 keV. In (b), energy interval varies from 0.01 keV, 0.1 keV, 1 keV, until 10 keV to show the energy distribution of PKAs.

The broad energy range of PKAs makes it challenging to simulate cascades for each PKA. Instead, we choose representative PKA energies which can both capture the major characteristics of cascades caused by different PKAs and make computational workload reasonable. PKAs of similar energies are grouped together and a representative energy within the energy group is used for further cascade simulations. Different energy intervals are applied along the energy axis, i.e., a 10 eV interval is used for energies lower than 100 eV, a 100 eV interval for energies lower than 1 keV, etc. In this approach, PKAs in each group will not exceed 20% of the total PKAs, while the number of necessary cascade simulations remains computationally accessible. The final results are shown in Figure 2 (b). In the low energy range (<100 eV), the number of PKAs first increases and then decreases with the PKA. In the higher energy range (>100 eV), the number of PKA continues to decrease with the increasing PKA energy. One should note that the apparent increase in the number of PKAs as well as peaks in Figure 2(b) are the result of an increasing energy interval used in the plot with increasing PKA energy.

## 3.2 MD
### 3.2.1 Cluster size distribution

The number and the size distribution of intra-cascade clusters produced by PKAs with given energies are averaged over six directions of the PKA velocity. The size distribution of clusters is shown in Figure 3. Clusters with more than 6 vacancies and with more than 4 interstitials are rare and the analysis in this paragraph will focus on clusters that are prevalent, which is $n \leq 6$ vacancies and $n \leq 4$ interstitials. Figure 3 shows that size distributions of both, interstitial and vacancy clusters, are shifted to larger sizes as the PKA damage energy $E_d$ increases from 0.01 to 50 keV. In addition, for each damage energy $E_d$, the number of clusters decreases when the cluster size increases.

We compared our result to those previously reported for specific cases, such as 30 keV Si PKA studied by MD [28]. In our study, 33 vacancy clusters of size 4 and 30 vacancy clusters of size 5 are found in cascades with PKAs distributed among six crystallographic directions. These numbers are significantly larger than those reported in Ref. [28], where cascade simulations with 30 keV PKAs distributed among three crystallographic directions were conducted and only one vacancy cluster of size 4 and two clusters of size 5 were found. In addition, in Ref. [28], the authors did not report any interstitial clusters containing more than 3 interstitials. In contrast, in our study we found 18 interstitial clusters of size 3, 9 interstitial clusters of size 4, 4 interstitial clusters of size 5, and one interstitial cluster of size 6. The most likely reason for the discrepancy is the criterion used to identify PDs clusters. In Ref. [28], the authors used a criterion based on the first nearest-neighbor (NN) distance (1.87 Å). This criterion is not sufficient to identify some of the defect clusters found by genetic algorithms [8] and DFT [9], where there is a finite binding energy of point defects separated by a distance larger than the first NN distance. Here, we adopt the 2nd NN as the criterion of defect clustering and we validate it by ensuring that this criterion can identify defects previously reported in Refs. [8] and [9]. 2nd NN is also commonly used in similar studies in metallic systems [19,45,46]. One should also note that our study and the one in Ref. [28] used different semi-empirical potentials in MD simulations. However, the effect of the potential is likely to be small, as both studies reported a larger number of interstitials and vacancies being generated for all PKAs and the main issue seems to be how to identify which defects are clustered.

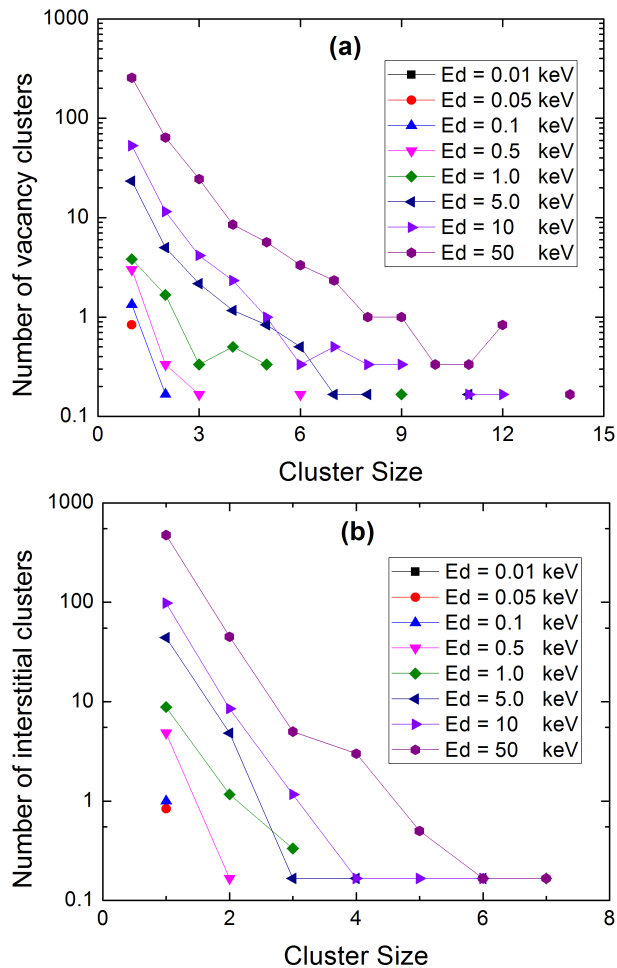

**Figure 3** Number of clusters produced by Si PKAs of certain damage energies $E_d$ as a function of damage energy $E_d$ and cluster size. (a) Production of vacancy clusters; (b) production of interstitial clusters. Lines are added to guide the eye.

Similarly, to what has been found in a number of metallic systems [18,19,22,47], size distributions of intra-cascade clusters produced by PKAs in SiC can be fitted to power law functions, $f \sim A/n^S$. $A$ and $S$ are fitting parameters that depend on the damage energy of the PKA and they are listed in Table 4. As the PKA damage energy increases, *A* increases, which means that more clusters are generated. Compared with intra-cascade vacancy cluster in tungsten [18,48], *S* is larger in SiC, which means that cascade gives a greater weight to smaller vacancy clusters. This trend might be due to a different qualitative dependence of binding energy of vacancy clusters on the cluster size in the two materials. In pure metals, binding energy of vacancy clusters generally increases with increasing cluster size[49,50], whereas in SiC binding energy of vacancy clusters decreases with increasing cluster size [51].

Intra-cascade clusters can act as nuclei for growth of larger clusters by attracting isolated PD or smaller clusters at the early stages of annealing [52]. Therefore, although there are only few interstitial clusters larger than 4 and vacancy clusters larger than 6, it is useful to know what is the largest size of a cluster that can be formed at a given PKA damage energy. As shown in Figures 4(a) and 4(b), the largest cluster size varies with the PKA direction and there is no specific direction that consistently leads to formation of the largest clusters across different PKA damage energies. The largest cluster size averaged over six directions is plotted in Figure 4 (c) to show a general trend. In most cases, the maximum size of vacancy clusters is larger than that of interstitial clusters. This trend can be qualitatively explained by the common observation that in irradiated materials vacancies remain close to the PKA path whereas interstitials are scattered away from it [19,35,53]. Specifically, during the collision phase of a cascade, PKAs displace a number of energetic atoms along the cascade path. Those energetic atoms dissipate their energies either by replacement collisions with neighbors or by moving away from the path of the PKA into the surrounding crystal. In both cases, vacancies are left near the PKA path to form a depleted zone and they can aggregate to form large clusters. Interstitials are scattered outside of the depleted zone and in the case of SiC they are not able to form large clusters.

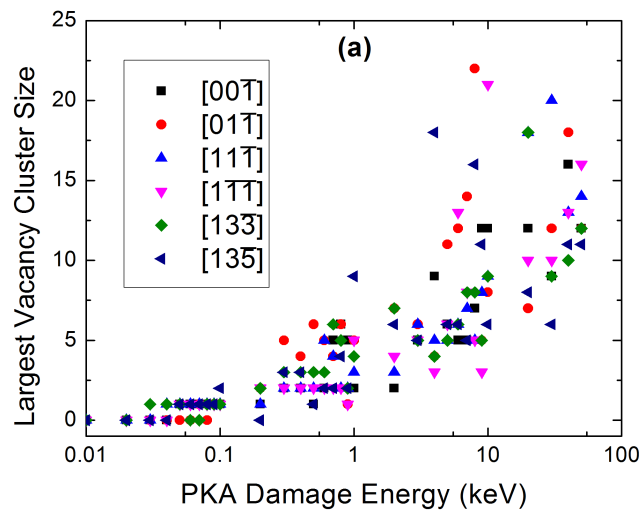

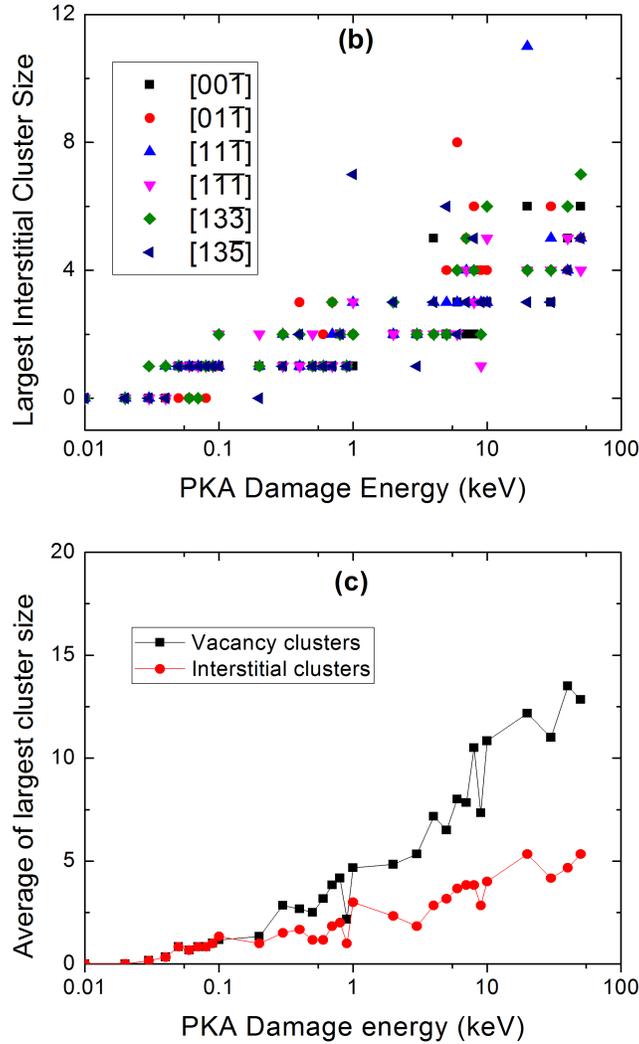

**Figure 4** The largest intra-cascade cluster size as a function of PKA damage energies and direction for (a) vacancy and (b) interstitial clusters. (c) shows a comparison of the average value of the largest cluster size between the six PKA directions. These directions are $[00\bar{1}]$, $[01\bar{1}]$, $[11\bar{1}]$, $[1\bar{1}\bar{1}]$, $[13\bar{3}]$ and $[13\bar{5}]$.

Interestingly, in Refs. [19,53] it was reported that iron and tungsten interstitials formed larger clusters than those formed by vacancies, despite the fact that interstitials were scattered outside the cascade path and vacancies remained close to the path. The main reason for this is the difference in cascade morphology between typical metal and SiC. In metals, cascade morphology is often characterized by a spherical vacancy-rich core surrounded by an interstitial-rich region[54], whereas in SiC cascades have a more elongated shape where interstitials are separated by the depleted zones and therefore cannot cluster easily[28]. In addition, localized melting in the cascade, which occurs in metals, does not take place in SiC. Instead, in SiC the cascade consists of isolated subcascades that branch off from the trajectory of the PKA, preventing development of compact cascades and limiting the number and size of defect clusters [55].

It is interesting to note that the largest cluster sizes among all cascade simulations are not found in simulations with the largest PKA damage energies of 50 keV. Specifically, Figure 4 (a) shows the largest vacancy cluster size is 22, which is produced by a PKA of 8 keV damage energy along the $[01\bar{1}]$ direction. Finally, as shown in Figure 4 (b), the largest interstitial cluster size is 11 and it corresponds to the damage energy of 20 keV along the $[11\bar{1}]$ direction. A possible reason for this phenomenon is that high energy cascades form isolated distributed subcascades [21,27], which may lower the possibility of forming large clusters. That also means that there is limit on how large a cluster can be formed directly in the cascade and that the frequency of producing such clusters is relatively small.

### 3.2.2 Cluster composition distribution

Figure 5 shows how carbon monomer concentrations in vacancy clusters and interstitial clusters depend on the cluster size. Hereafter, monomer refers to a vacancy or an interstitial when vacancy clusters and interstitial clusters are discussed, respectively. Figure 5 shows that carbon vacancy concentration in vacancy clusters varies from 30% to 100%, and carbon interstitial concentration in interstitial clusters has a narrower range, which is from 60% to 100%. The majority of vacancy clusters are made of 70% - 90% carbon vacancies, and most interstitial clusters consist of 80% - 100% carbon interstitials. Among all vacancy clusters, 63.9% are carbon rich, 32.1% are silicon rich, and 3.97% are stoichiometric. Among all interstitial clusters, the corresponding percentages are 86.5%, 6.3%, and 7.17%. The reason why there are more carbon than silicon monomers in both interstitial and vacancy clusters is easily understood, because threshold displacement energies of carbon and silicon are 20 eV and 35 eV, respectively [28,31–33] which means that it is easier to displace carbon than silicon atoms during a cascade [29,37]. The observation that concentration of carbon monomers in interstitial clusters is higher than in vacancy clusters can be rationalized as follows. During a cascade, interstitials and vacancies are formed either by displacement or by replacement. During displacement events, the numbers of interstitials and vacancies produced are the same, and this process leads to equal concentrations of carbon monomers in interstitial and in vacancy clusters. The situation is somewhat different for replacement events. If replacement takes place between the same species, such as a carbon interstitial and a carbon atom, there will be no change in composition ratios. On the other hand, if replacement occurs between different species, such as a carbon interstitial and a silicon atom, the composition ratios will change. The difference between threshold displacement energies of carbon and silicon makes it easier for silicon interstitials to replace carbon atoms, which results in a higher concentration of carbon monomers in interstitial clusters than in vacancy clusters.

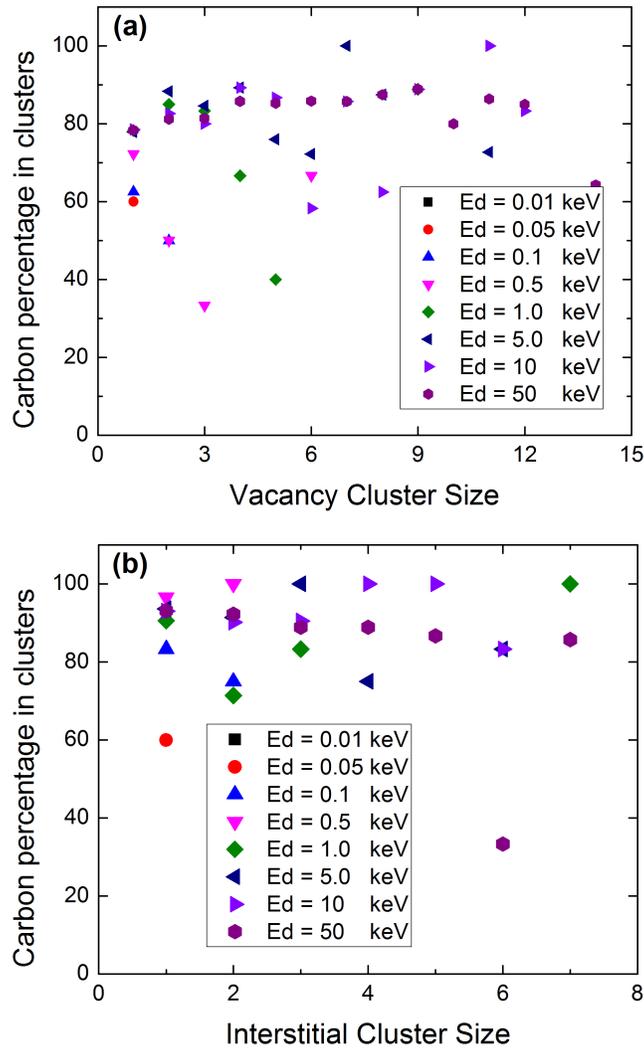

**Figure 5** Percentage of carbon in clusters created by PKA with different energies as a function of cluster size. Cluster size is defined as the number of defects in the cluster. $E_d$ stands for damage energy. (a) percentage of carbon vacancies in vacancy clusters; (b) percentage of carbon interstitials in interstitial clusters.

The trends observed in our simulations are consistent with previously published data. For example, in cascade simulations with 4 and 10 keV Si PKAs reported in Refs. [16,36], the number of silicon antisites ($Si_C$) was found to be almost twice as large as the number of carbon antisites ($C_{Si}$), the ratio between the number of carbon interstitials and the number of silicon interstitials was 10.86:1, and the ratio between the number of carbon vacancies and the number of silicon vacancies was 3.92:1. As clusters are formed by adjacent vacancies or interstitials, it follows that if the point defects reported in Refs. [16,36] formed clusters, the concentration of carbon monomers in interstitial clusters would be higher than that in vacancy clusters, which is exactly the finding in our current study.

It has been speculated by Katoh *et al.* [56] that significantly large clusters, which form as a result of long-term evolution of radiation-induced defects, maintain a stoichiometric structure. This hypothesis was verified by atomistic simulations of Ko *et al.* [30] and Watanabe *et al.* [8] who found that for clusters larger than 10 point defects, stoichiometric clusters are more stable than off-stoichiometric clusters. Based on this information, cluster dynamics models [17] assumed that intra-cascade clusters (including small clusters) in SiC are stoichiometric and that clusters can grow by absorbing the same number of carbon and silicon atoms. In Ref. [17] the simulated clusters were found to grow significantly faster and to larger sizes than those observed in experiments. Here, we find that intra-cascade clusters are carbon rich and that stoichiometric clusters constitute only a small fraction of clusters produced directly in the cascade. It is possible that the C-rich clusters produced in cascades preferentially absorb Si interstitials in order to achieve the desired 1:1 stoichiometry. As Si interstitials have a larger migration barrier than C interstitials and are produced in smaller quantities, cluster growth by absorbing preferentially Si interstitials would be slower than by absorbing both Si and C atoms. It needs to be verified whether including the possibility of off-stoichiometric intra-cascade clusters in cluster dynamics models like that from Ref. [17] would lead to the growth rate slow enough to be comparable to experiments..

### 3.3 Integration

One PKA from each energy interval in the PKA energy spectrum is used to conduct MD simulations of cascades, and we assume that the result can represent the damages caused by other PKAs in the same group. For example, we use the 10 eV MD cascade simulation result to represent the damage caused by all PKAs in the energy interval of 0 eV – 10 eV, which provides the upper bound on the cascade efficiency. C and Si PKAs that have damage energy larger than 50 keV are neglected due to their low percentages, i.e., 0.0005% and 0.177% for C and Si, respectively.

The numbers of Frenkel pairs (FPs) produced by PKAs of certain damage energy averaged over six PKA directions are multiplied by the production rate of PKAs in the corresponding energy interval (energy intervals are defined in Section 3.1). The results are shown in Figure 6, where we plot the number of FPs generated by all PKAs in a given energy interval. It is known that the number of FPs generated by PKA of a certain energy increases linearly as the PKA damage energy increases [28]. Although the fractions of PKAs with large energies are small as compared to those with small energies (Figure 2), as the number of FPs introduced by high-energy PKAs is large, the total contributions from high-energy PKAs to the final FP count are dominant (Figure 6). In addition, the majority (i.e., 84.9%) of surviving FPs are generated by PKAs with damage energy larger than 1 keV. Carbon PKAs created more FPs because the damages caused by C and Si with the same energies are similar, but there are more C PKAs in the system.

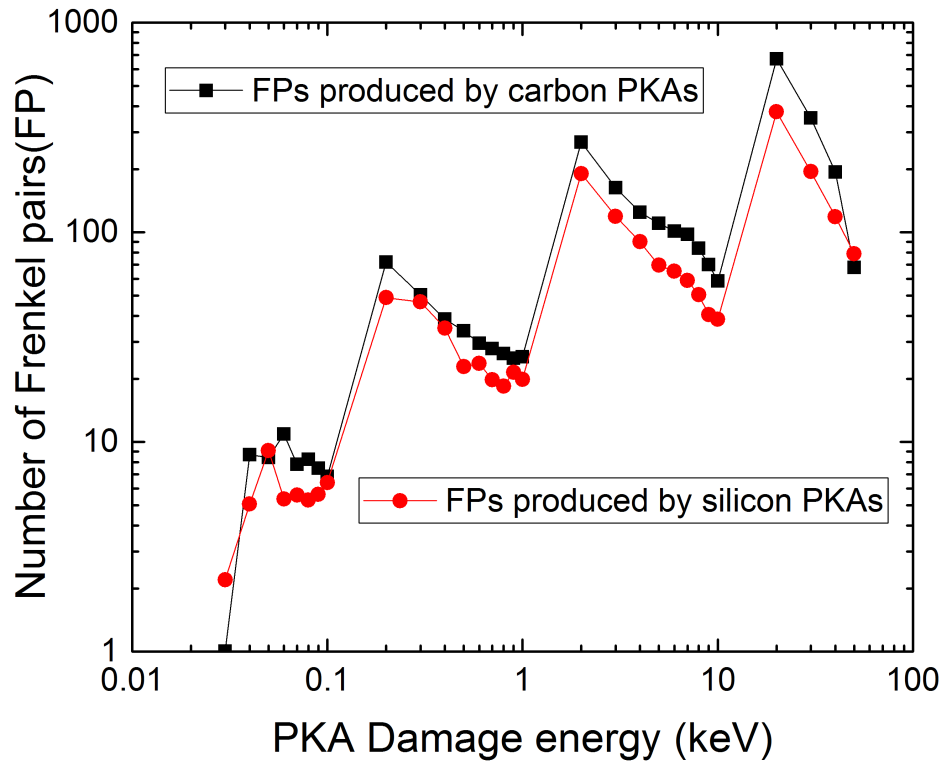

**Figure 6** Number of Frenkel pairs produced by PKAs with a given damage energy.

Similarly to the FP analysis, the numbers of intra-cascade clusters of size $n$ produced by PKAs with a certain damage energy are averaged over six PKA directions. These average values are then multiplied by the production rate of corresponding PKAs to obtain the production rate of intra-cascade cluster of size $n$ by all PKAs in the certain damage energy interval. Next, for each cluster size $n$, we add up the numbers of clusters produced in each energy interval to determine the production rate of cluster of size $n$ by one incoming ion (in our example 1 MeV Kr). These values are further divided by the total number of FPs generated by one 1 MeV Kr ion and by the size of cluster $n$ to determine a function that represents the production of vacancy and interstitial clusters per a successful displacement. The results are shown in Figure 7 (a). The production functions for both vacancy and interstitial clusters per successful displacement, $F$, are fitted to a power law $F \sim A/n^S$ with the fitted values being $A_V$ = 0.429 ± 0.011, $S_V$ = 2.092 ± 0.114 and $A_I$ = 0.804 ± 0.002 and

$S_I$ = 3.591 ± 0.005. The value of $S$ found here for SiC is significantly larger than those determined from cascade simulations in metallic systems [18,19,22]. The value of $S$ represents the tendency of forming smaller clusters, which means that in SiC small intra-cascade clusters are more preferable than in metallic systems. Figure 7 (a) also shows that compared with vacancies, interstitials are more likely to remain isolated from each other. Specifically, more than 80.4% of interstitials are single interstitials, whereas only 44.9% of vacancies are isolated from each other. Also, even if interstitials form intra-cascade clusters, they prefer to form smaller clusters than vacancies, which is further evidenced by the larger size of the largest interstitial clusters as compared to the largest vacancy cluster shown in Figure 4 and by the slope between the FP production function and the clusters size shown in Figure 7 (a).

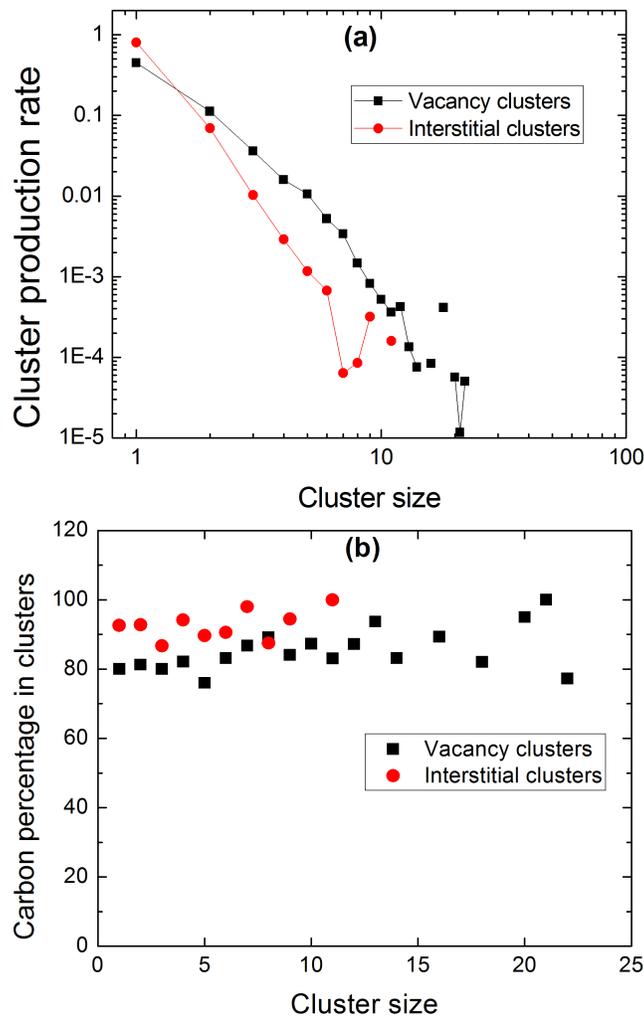

**Figure 7** (a) Intra-cascade cluster production function per successful displacement caused by 1 MeV Kr ion in 3C-SiC. (b) Carbon percentage in vacancy and interstitial clusters produced by 1 MeV Kr ion in 3C-SiC.

Using a similar approach, carbon monomer concentrations in vacancy and interstitial clusters are analyzed and shown in Figure 7 (b). It is clear that carbon monomer is the dominant species in both, vacancy and interstitial clusters. The average carbon monomer composition for interstitial clusters is 90.1%, which is higher than the composition in vacancy clusters, which is 84.6%. The reasons for this trend are the same as those for trends reported in Figure 5 and they have been already discussed in Section 3.2.2.

The cascade efficiency of PKA is defined as the ratio of successful displacements (SDs) at the end of the cascade over the number of total displacements (TDs) generated during cascade. TD has been often estimated using Norgett, Robinson, and Torrens (NRT) model [28,34,57]:

$$N_{NRT} = \frac{0.8 E_{PKA}}{2 E_{th}},$$

where $E_{th}$ is the average threshold displacement energy, and $N_{NRT}$ is an estimate of TD. The threshold displacement energy for carbon is 20 eV and for silicon is 35 eV[28,31–33], hence $E_{th}$ should lie between 20 eV and 35 eV. However, it has been also argued that the NRT model is not an accurate estimate of TDs [57]. A more accurate estimate of TDs is to count TDs at the thermal spike of cascade. In SiC the number of displacements generally reaches a peak between 0.1 ps to 0.3 ps of the collision cascade [39] and remains approximately constant after 10ps. Here, we estimated the value of TDs at 0.1 ps, 0.2 ps, 0.3 ps, and 0.4 ps and we chose the largest number among these values. The total displacements in our analysis are determined as the number of defects, including vacancies, interstitials, and antisites, regardless of whether they are isolated or part of clusters. Successful displacements are determined as that total number of defects that survived from cascade thermal spike and quenching and are still present in the system at 12 ps of the cascade (where these defects diffuse very slowly and the number of defects reaches an approximately constant value on the time scales of MD simulations[11]). As an example, cascade efficiencies for Si PKA with the velocity vector along the [1$\overline{1}\overline{1}$] direction with various energies are shown in Figure 8. The efficiency is equal to 0 when the PKA damage energy is smaller than 40 eV, which is less than or close to the threshold energy of Si PKA. Displaced atoms in these cascades will quickly recombine with vacancies left on the lattice. For energies from 50 eV to 50 keV, cascade efficiency fluctuates and decreases from 1 at 0.05 keV to 0.6 in the range of 10 – 50 keV. Compared with the conclusion in Ref. [28], where cascade efficiency decreases from 1 at 0.25 keV to 0.5 at 30 – 50 keV, here we extend the lower limit of damage energies to 0.05 keV for which cascade efficiency are 1.

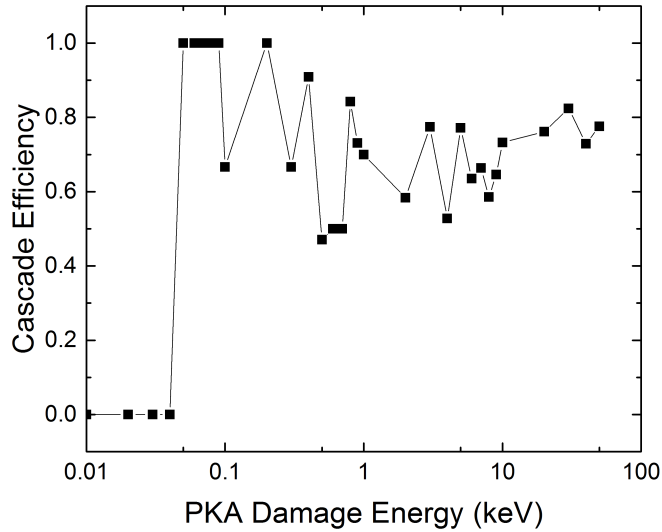

**Figure 8** Cascade efficiency as a function of Si PKA damage energy with the PKA velocity along the $[1\bar{1}\bar{1}]$ direction.

## 4. Conclusion

We have conducted MD simulations of cascades caused by PKAs with energies from 0.01 keV to 50 keV with velocities along the following directions $[00\bar{1}]$, $[01\bar{1}]$, $[11\bar{1}]$, $[1\bar{1}\bar{1}]$, $[13\bar{3}]$ and $[13\bar{5}]$ in order to determine the distributions of intra-cascade cluster size and composition in SiC. Both intra-cascade interstitial and vacancy clusters productions obey a power law and vacancy clusters are found to be larger than interstitial clusters. Carbon is the dominant species in both interstitial and vacancy clusters and stoichiometric clusters constitute only 7.4% and 3.9% of all the interstitial and vacancy clusters, respectively. Carbon vacancy ratio in vacancy clusters is lower than C interstitial ratio in interstitial clusters, which can be explained by the fact that the production rate of $Si_C$ is higher than $C_{Si}$. The calculated distributions for intra-cascade clusters can be integrated with the PKA energy spectrum from ion irradiations to determine the overall production function of intra-cascade clusters and their compositions. 1 MeV Kr ion irradiated 3C-SiC is used as an example in the current study. Both vacancy and interstitial clusters obey power law $f = A/n^S$ with $A_V$ = 0.429 ± 0.011, $S_V$ = 2.092 ± 0.114 and $A_I$ = 0.804 ± 0.002 and $S_I$ = 3.591 ± 0.005. The average carbon monomer ratios are 91.5% and 85.3% for interstitial and vacancy clusters, respectively.

**Acknowledgements**

The authors gratefully acknowledge financial support from the US Department of Energy Basic Energy Science Grant # DE-FG02-08ER46493. We also acknowledge helpful discussions with Prof. Dane Morgan from UW-Madison and Dr. Xing Wang from ORNL.